\documentstyle[twoside,fleqn,espcrc2]{article}
\input epsf

\newcommand{\AmS}{{\protect\the\textfont2
  A\kern-.1667em\lower.5ex\hbox{M}\kern-.125emS}}

\title{Finite Size Effects in Quark-Gluon Plasma Formation }

\author{Andy Gopie and Michael C. Ogilvie
\address{Department of Physics,
        Washington University, \\
        St. Louis, MO 63130 USA}%
        \thanks{We thank the U.S. Department of Energy
        for financial support under
        grant number DE-FG02-91-ER40628.}
}
       
\begin{document}

\begin{abstract}
Using lattice simulations of quenched QCD
we estimate the finite size effects
present when a gluon plasma equilibrates in a slab geometry,
{\it i.e.}, finite width but large transverse dimensions.
Significant differences are observed in the free energy density for the
slab when compared with bulk behavior. A small shift in the critical
temperature is also seen. The free energy required to liberate
heavy quarks relative to bulk is measured using
Polyakov loops; the additional free energy required is
on the order of $30-40 ~MeV$ at $2-3~T_c$.
\end{abstract}

\maketitle


Lattice gauge theory has given us considerable
information about bulk
thermodynamics, but finite size effects have up to now been
studied using simplified, phenomenological models.
We study the behavior of a gluon plasma restricted to a slab geometry, with
the longitudinal width much smaller than the transverse directions.
This inner region is heated to temperatures above the bulk deconfinement
temperature, surrounded by an outer region which is kept at a temperature
below the deconfinement temperature,
providing confining boundary conditions for the
inner region.
Details of this work are given in \cite{Gopie}.

By allowing the coupling constant $\beta = 6/g^2$ to
vary, a spatially dependent temperature
can be introduced.
We have chosen the temperature interface to be sharp,
in such a way that the lattice is
divided into two spatial regions, one hotter and one colder.
The quenched approximation simplifies the role
of the cold region, because below $T_c$,
the dominant excitation at low energies is the scalar glueball.
Temperatures near $T_c$ are 
smaller than glueball masses by about a
factor of four, 
so glueballs play no 
the pressure in the hadronic phase is essentially zero.
We thus expect the slab thermodynamics to be largely
insensitive to the precise temperature of the region outside the
slab, as long as it is sufficiently low.
In full QCD, this insensitivity to the outer temperature
would not hold, due to pions.
Note that the role of boundary conditions here is quite
different from that in bubble nucleation. In that case, both
$\beta_{in}$ and $\beta_{out}$ are taken to be near $T_c$.
\cite{Kajantie}\cite{Huang2}

The slab is given a
fixed lattice width, $w=6a$, rather than of fixed physical width.
Since $N_t = 4$, $w T$ is fixed at $3/2$. 
At higher temperatures, the width in physical units
is somewhat smaller than the
longitudinal size of the plasma formation region expected in heavy
ion collisions.
While the use of equilibrium statistical mechanics to study gluon plasma
properties during the early stages of plasma formation may appear
suspect, a simple estimate using the Bjorken model \cite{Bjorken}
shows that when a coin-shaped region of
width 1 fermi has expanded to 1.5 fermi, the variation in temperature
is only from $0.8 T_0$ at the center of the coin to $T_0$ at its edges.

The free energy density $f$ for the slab was obtained using the standard
method \cite{Engels} of integrating the lattice action with respect to
$\beta$. We use a convenient convention for the sign of $f$
that is opposite the usual one. In the bulk case, $f$ is then identical
to the pressure $p$.

\begin{eqnarray}
{f \over T^4}|_{\beta_{out}}^{\beta} = 
N_t^4 \int_{\beta_{out}}^{\beta} d\beta' 
\left[ \langle S \rangle_T - \langle S \rangle_0 \right]
\end{eqnarray}
where $ S  =  (1/N) Re Tr\ U_p$.
As in the bulk case, it is necessary to subtract the
zero-temperature expectation value from the finite temperature
expectation value, in this case
using the same pair of $\beta$ values.
In general, quantum field theories with boundaries develop divergences that
are not present in infinite volume or with periodic boundary conditions.
It is expected that any counterterms would be
gauge-invariant and have dimension $ \leq 3$; there are no such
counterterms.\cite{Symanzik} 

In order to take advantage of the data on bulk thermodynamics
provided by the Bielefeld group \cite{Boyd}, we worked consistently
with lattices of overall size $16^3 \times 4$.
The values used for each subtraction come from
$16^4$ lattices with identical values of $\beta_{in}$ and
$\beta_{out}$.
The value of $\beta_{out}$ was held fixed at $5.6$ while
$\beta_{in}$ varied from $5.6$ to $6.3$.
For comparison, the bulk transition for $N_t = 4$ occurs at
$\beta_c (N_t=4, N_s=16) = 5.6908 (2)$
$\beta_c (N_t=4, N_s=\infty) = 5.6925 (5)$.
\cite{Boyd}


Figure 1 shows the free energy density $f/T^4$ versus $T/T_c$ compared with
the bulk pressure.
The free energy in the slab is lower than the bulk value by almost
a factor of two at $2 T_c$. It appears that the slab value is
slowly approaching the bulk value, but other behaviors are also
possible. 
Calculations of the finite-temperature contribution to the
Casimir effect for a free Bose field contained between two plates
show that $f/T^4$ has a non-trivial dependence on the dimensionless
combination $w T$.\cite{Mehra} \cite{Plunien}
It is natural to ask if the corrections to the free energy seen here
can be accounted for by the conventional Casimir effect.
A straigtforward calculation of the free energy of a
non-interacting gluon gas confined to a slab shows an increase in
the free energy density over the bulk value by a factor of about 1.63
at $wT = 3/2$. The Casimir effect alone is thus unable to explain the
reduction of the free energy observed in our simulations.
We are currently investigating more elaborate theoretical models which
include the effect of a non-trivial Polyakov loop.

\begin{figure}[htb]
\epsfxsize=75mm \epsfbox{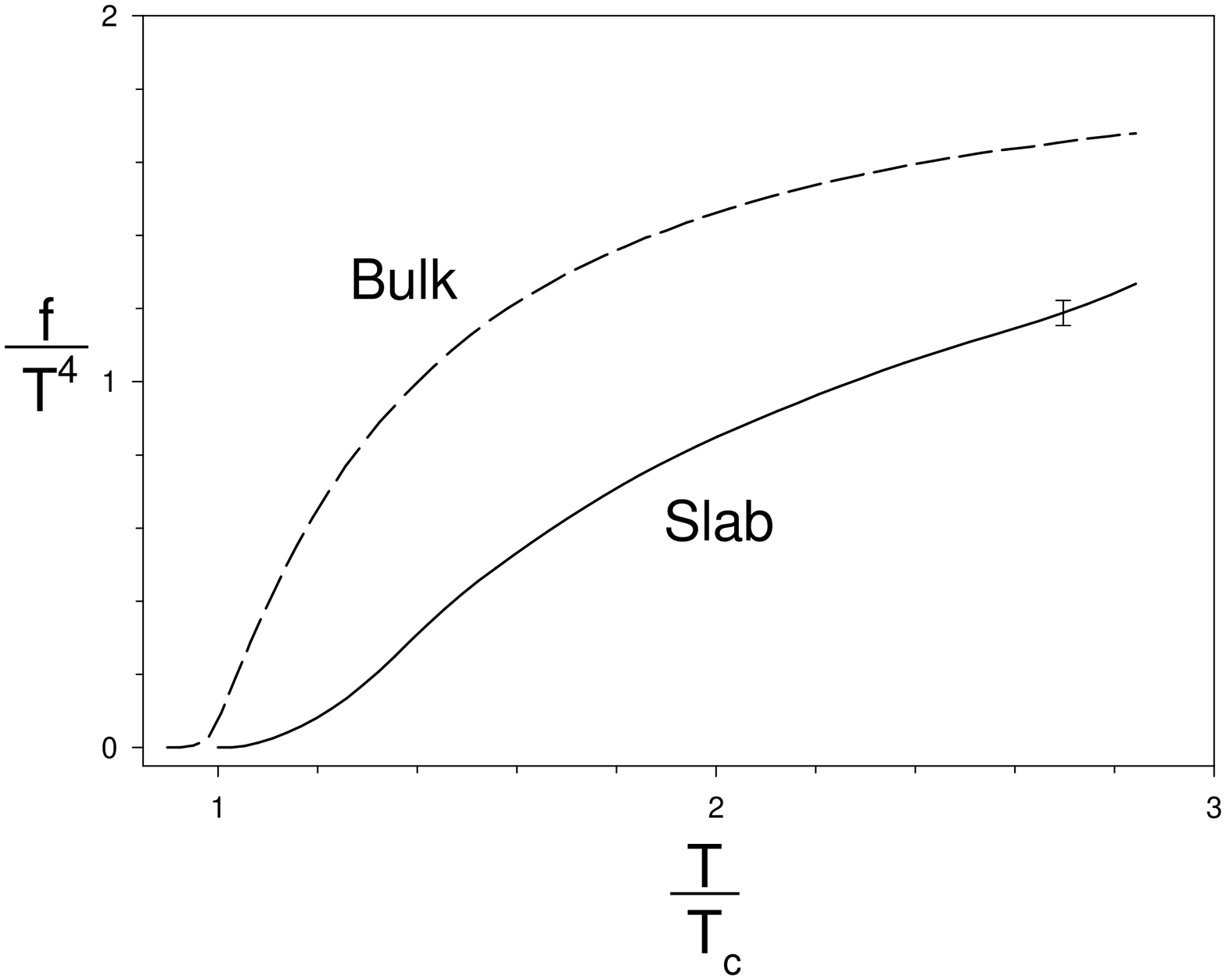}
\vspace{-0.3in}
\caption{Slab free energy compared with bulk pressure.}
\label{fig:fig1}
\end{figure}

A consistency check was performed on the surface effects.\cite{Huang}
The free energy density was calculated for a system at $\beta_{in} = 6.0$
by performing simulations with
$\beta_{in}$ fixed at $6.0$ and $\beta_{out}$ varying from $5.6$ to $6.0$.
Combining these results with the bulk data of reference \cite{Boyd}
creates a path equivalent to varying $\beta_{in}$ while holding
$\beta_{out}$ fixed.
For $\beta_{in} = 6.0$ and $\beta_{out}=5.6$, 
this gives $f/T^4 = 0.65 \pm 0.04$,
to be compared with $f/T^4 = 0.69 \pm 0.03$
for the direct calculation.

The major source of systematic error lies with the choice of
boundary conditions for the slab, here set by $\beta_{out}$.
We have estimated the 
systematic error associated with the choice of boundary conditions by
by performing simulations at $\beta_{in} = 6.2$
and $\beta_{out} = 5.5$ on $16^3 \times 4$ and $16^4$ lattices.
These results suggest that lowering $\beta_{out}$
from $5.6$ to $5.5$ reduces the free energy by roughly
10 percent at $\beta_{in} = 6.2$.


We define an effective surface tension $\alpha(w,T)$ by
\begin{eqnarray}
f = p - 2 \alpha(w,T) / w
\end{eqnarray}
where the notation $\alpha(w,T)$ recognizes that the surface tension
$\alpha$ does depend on the width of the slab and the internal
and external temperatures.
In the limits where $T$ approaches $T_c$ and $w$ goes to infinity,
this quantity approaches $\alpha_0$.
Figure 2 shows $\alpha(w,T)/T^3$ versus $T/T_c$ for $w T = 3/2$;
representative error bars are shown.
Away from the bulk critical point, $\alpha / T^3$ rises quickly
to a peak at about $1.4 T_c$, and then falls slowly as $T$ increases.
A large non-equilibrium surface tension has also been observed
in measurements of the equilibrium surface tension $\alpha_0$,
where these
effects were obstacles to obtaining $\alpha_0$.\cite{Huang}

\begin{figure}[htb]
\epsfxsize=75mm \epsfbox{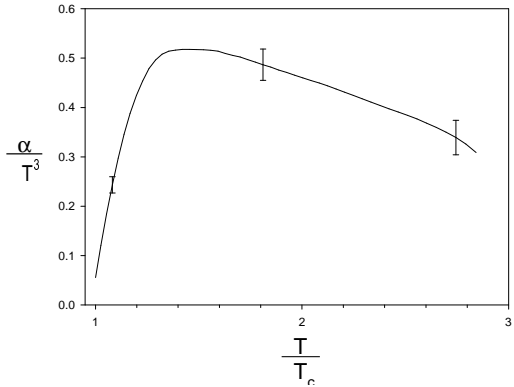}
\vspace{-0.3in}
\caption{Non-equilibrium surface tension versus $T$.}
\label{fig:fig2}
\end{figure}

The multiplicative divergence in the Polyakov loop 
can be eliminated when comparing
bulk expectation values to those in finite geometries.
We define
\begin{eqnarray}
\Delta F_Q(\vec{x}) =  -T~ ln \left[ P_{slab}(\vec{x})/P_{bulk} \right]
\end{eqnarray}
as the excess free energy required to liberate a heavy quark
in the slab geometry relative to bulk quark matter at the
same temperature. 

In figure 3, we show the 
expectation value for the Polyakov loop versus
z measured in lattice units for several values of $\beta$.
Each curve is normalized by dividing the values of the Polyakov
loop by the bulk expectation value at the corresponding value of
$\beta$.
Error bars are shown only for even values of z.
It is clear that a significant change occurs between 
$\beta = 5.8 ~~~(T= 1.23~T_c)$ and $\beta = 5.85 ~~~(T= 1.36~T_c)$.
For larger values of $\beta$, $\Delta F_Q$ diminishes to a
value  of approximately $30 - 40 ~~MeV$ in the middle of the slab.

There are significant deviations in the
slab geometry from bulk behavior and the ideal gas law,
arising from a strong non-equilibrium
surface-tension. This non-equilibrium surface tension can be
an order of magnitude greater
than the equilibrium value.
Surface tension effects also produce a
mild elevation of the apparent critical temperature.
Measurement of Polyakov loop expectation values relative to bulk
shows that the suppression of heavy quark production
due to the slab geometry is small.

There are good reasons to call our result an estimate rather than
a calculation. Although lattice gauge theory simulations of bulk
behavior can be made arbitrarily accurate in principle, in this case
there is uncertainty in the exterior boundary conditions
appropriate; the applicability of equilibrium
thermodynamics at this early stage of quark-gluon plasma formation
can also be questioned.
However, this result is the best estimate available now,
and further refinements are possible.

\vspace{-0.3in}
\begin{figure}[htb]
\epsfxsize=75mm \epsfbox{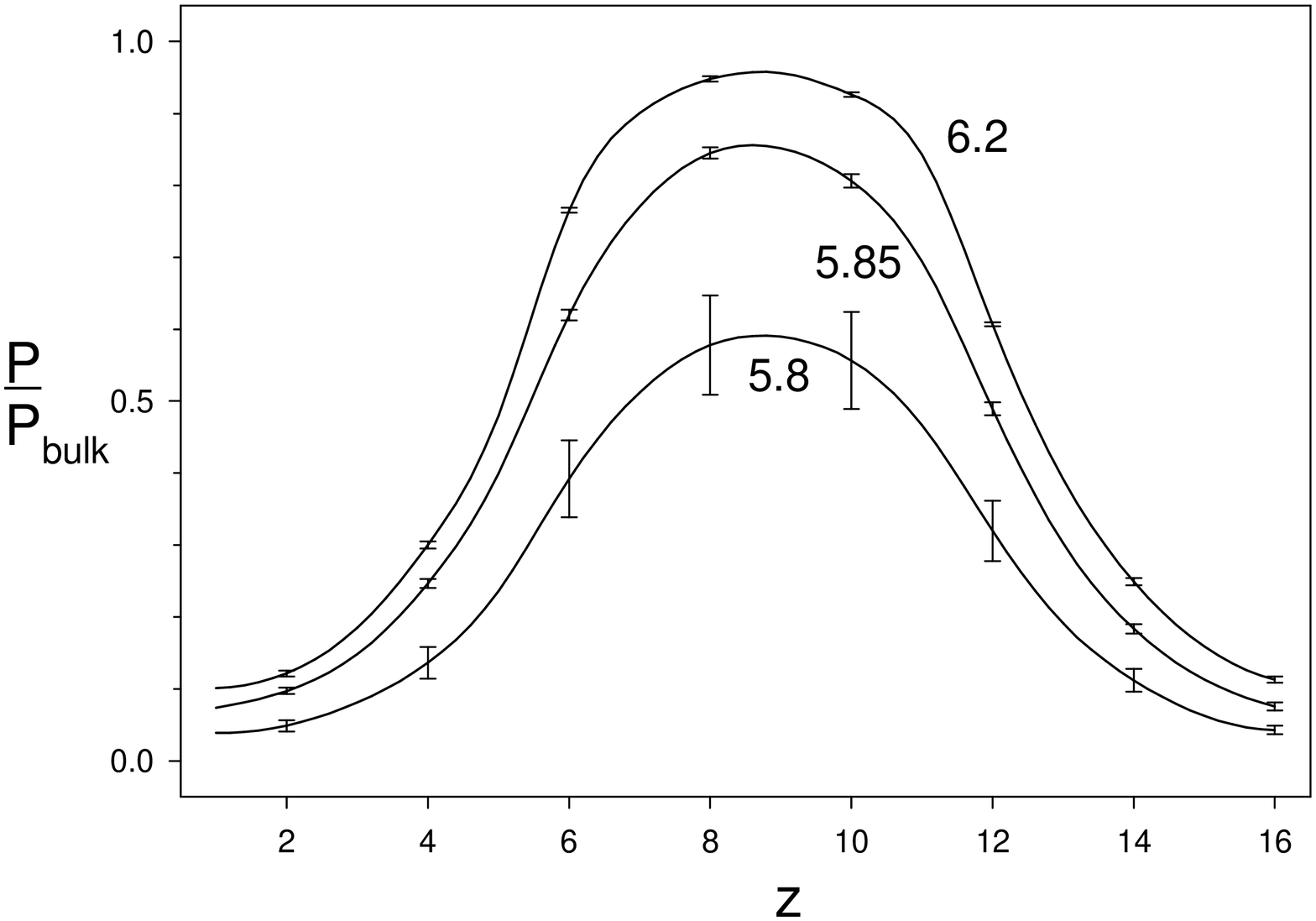}
\vspace{-0.3in}
\caption{Polyakov loop versus $z$.}
\label{fig:fig3}
\end{figure}

\end{document}